\documentclass[conference]{IEEEtran}
\pdfoutput=1

\usepackage{bm}
\usepackage{graphicx}
\usepackage{booktabs}

\usepackage[per-mode=symbol]{siunitx}    		
\DeclareSIUnit{\dBm}{dBm}	

\usepackage{subcaption}
\usepackage[
indexonlyfirst, 
]{glossaries}
\newacronym{aoa}{AOA}{angle-of-arrival}
\newacronym{aod}{AOD}{angle-of-departure}
\newacronym{asic}{ASIC}{application specific integrated circuit}
\newacronym{awgn}{AWGN}{additive white Gaussian noise}
\newacronym{bs}{BS}{base station}
\newacronym{ccs}{CCS}{correlative channel sounder}
\newacronym{ce}{CE}{channel estimation}
\newacronym{csi}{CSI}{channel state information}
\newacronym{csp}{CSP}{Contact Service Point}
\newacronym{crlb}{CRLB}{Cram\'er-Rao lower bound}
\newacronym{da}{DA}{data association}
\newacronym{dc}{DC}{direct current}
\newacronym{dm}{DM}{diffuse multipath}
\newacronym{dmc}{DMC}{\gls{dm} component}
\newacronym{ecsp}{ECSP}{Edge Computing Service Point}
\newacronym{eirp}{EIRP}{equivalent isotropically radiated power}
\newacronym{en}{EN}{energy neutral}
\newacronym{end}{END}{energy neutral device}
\newacronym{epu}{EPU}{edge processing unit}
\newacronym{fa}{FA}{federation anchor}
\newacronym{gbsc}{GBSC}{geometry-based deterministic stochastic channel}
\newacronym{gsm}{GSM}{Global System for Mobile Communications}  %
\newacronym{ic}{IC}{integrated circuit}
\newacronym{id}{ID}{information decoding}
\newacronym{iot}{IoT}{Internet of Things}
\newacronym{lis}{LIS}{large intelligent surface}
\newacronym{los}{LoS}{line-of-sight}
\newacronym{lsa}{LSA}{large synthetic array}
\newacronym{lti}{LTI}{linear time-invariant}
\newacronym{mmwave}{mmWave}{millimeter wave}
\newacronym{mimo}{MIMO}{multiple-input multiple-output}
\newacronym{miso}{MISO}{multiple-input single-output}
\newacronym{mpc}{MPC}{multipath component}
\newacronym{mrt}{MRT}{maximum ratio transmission}
\newacronym{nfc}{NFC}{near-field communication}
\newacronym{ofdm}{OFDM}{orthogonal frequency-division multiplexing}
\newacronym{ota}{OTA}{over-the-air}
\newacronym{nb}{NB}{narrowband}
\newacronym{nlos}{NLoS}{non-line-of-sight}
\newacronym{pa}{PA}{power amplifier}
\newacronym{pl}{PL}{path loss}
\newacronym{pc}{PC}{pilot count}
\newacronym{rcs}{RCS}{radar cross section}
\newacronym{rf}{RF}{radio frequency}
\newacronym{rfid}{RFID}{radio frequency identification}
\newacronym{ris}{RIS}{reflective intelligent surface}
\newacronym{rms}{RMS}{root mean square}
\newacronym{rss}{RSS}{received signal strength}
\newacronym{rssi}{RSSI}{received signal strength indicator}
\newacronym{ra}{RA}{reflectarray}
\newacronym{rw}{RW}{RadioWeaves}
\newacronym{sa}{SA}{synchronization anchor}
\newacronym{sar}{SAR}{specific absorption rate}
\newacronym{sbl}{SBL}{sparse Bayesian learning}
\newacronym{sdn}{SDN}{software-defined network}
\newacronym{sinr}{SINR}{signal-to-interference-plus-noise ratio}
\newacronym{simo}{SIMO}{single-input multiple-output}
\newacronym{siso}{SISO}{single-input single-output}
\newacronym{slam}{SLAM}{simultaneous localization and mapping}
\newacronym{snr}{SNR}{signal-to-noise ratio}
\newacronym{smc}{SMC}{specular multipath component}
\newacronym{s-parameters}{S-parameters}{scattering parameters}
\newacronym{tosm}{TOSM}{through-open-short-match}
\newacronym{tdoa}{TDOA}{time-difference-of-arrival}
\newacronym{toa}{TOA}{time-of-arrival}
\newacronym{ue}{UE}{user equipment}
\newacronym{uhf}{UHF}{ultra high frequency}
\newacronym{ula}{ULA}{uniform linear array}
\newacronym{upa}{UPA}{uniform planar array}
\newacronym{ura}{URA}{uniform rectangular array}
\newacronym{uwb}{UWB}{ultrawideband}
\newacronym{vna}{VNA}{vector network analyzer}
\newacronym{wb}{WB}{wideband}
\newacronym{wpt}{WPT}{wireless power transfer}
\newacronym{zf}{ZF}{zero forcing}

\newacronym{cdf}{CDF}{cumulative distribution function}
\newacronym{ssf}{SSF}{small-scale fading}
\newacronym{lsf}{LSF}{large-scale fading}
\newacronym{ple}{PLE}{pathloss exponent}
\newacronym{usrp}{USRP}{universal software radio peripheral}
\newacronym{rb}{Rb}{Rubidium}
\newacronym{pps}{1PPS}{1 pulse-per-second}
\newacronym{mrc}{MRC}{maximum ratio combinig}
\newacronym{fpga}{FPGA}{field-programmable gate array}
\newacronym{agc}{AGC}{automatic gain control}
\newacronym{tdma}{TDMA}{time-division multiple access}
\newacronym{cp}{CP}{cyclic prefix}
\newacronym{dac}{DAC}{digital-to-analog converter}

\newacronym{pdf}{PDF}{probability density function}
\newacronym{cbsm}{CBSM}{correlation-based stochastic model}
\newacronym{gbsm}{GBSM}{geometry-based stochastic model}
\newacronym{vr}{VR}{visibility region}
\newacronym{acf}{ACF}{autocorrelation function}

\newacronym{jcas}{JCAS}{joint communication and sensing}
\newacronym{v2x}{V2x}{vehicle-to-everything}
\newacronym{m2m}{M2M}{machine-type communication}
\newacronym{dmimo}{D-MIMO}{distributed MIMO}
\newacronym{urllc}{URLLC}{ultra-reliable and low-latency communication}%

\newcommand{\V}[1]{\ensuremath{\bm{#1}}}

\newlength{\belowFigureMargin}
\title{Large Intelligent Surface Measurements for\\%
Joint Communication and Sensing}%

\author{\IEEEauthorblockN{Christian Nelson\IEEEauthorrefmark{3},
Xuhong Li\IEEEauthorrefmark{3},
Thomas Wilding\IEEEauthorrefmark{4},
Benjamin Deutschmann\IEEEauthorrefmark{4}\\
Klaus Witrisal\IEEEauthorrefmark{4},
Fredrik Tufvesson\IEEEauthorrefmark{3}}

\IEEEauthorblockA{\IEEEauthorrefmark{3}Dept. of Electrical and  Information Technology, Lund University, Sweden}
\IEEEauthorblockA{\IEEEauthorrefmark{4}Signal Processing and Speech Communication Laboratory, Graz University of Technology, Austria}
 Email: \{chistian.nelson, xuhong.li, fredrik.tufvesson\}@eit.lth.se}
\begin{document}
\maketitle

\begin{abstract}
Multiple concepts for future generations of wireless communication standards utilize coherent processing of signals from many distributed antennas. Names for these concepts include distributed MIMO, cell-free massive MIMO, XL-MIMO, and large intelligent surfaces. They aim to improve communication reliability, capacity, as well as energy efficiency and provide possibilities for new applications through joint communication and sensing. One such recently proposed solution is the concept of RadioWeaves. It proposes a new radio infrastructure for distributed MIMO with distributed internal processing, storage and compute resources integrated into the infrastructure.
The large bandwidths available in the higher bands have inspired much work regarding sensing in the mmWave- and sub-THz-bands, however, sub-6\,GHz cellular bands will still be the main provider of broad cellular coverage due to the more favorable propagation conditions. In this paper, we present results from a sub-6\,GHz measurement campaign targeting the non-stationary spatial channel statistics for a large RadioWeave and the temporal non-stationarity in a dynamic scenario with RadioWeaves. From the results, we also predict the possibility of multi-static sensing and positioning of users in the environment.
\end{abstract}

\begin{IEEEkeywords}
channel characterization, channel measurement, distributed MIMO, dynamic, RadioWeaves,  sensing
\end{IEEEkeywords}

\section{Introduction}
\label{sec:Introduction}
With the advances of the fifth and sixth generation (5G and 6G) of mobile communication systems, new application fields are emerging; fields like vehicle-to-everything, machine-to-machine communication, and smart cities \cite{3gppURLLC}.
Furthermore, new frequency bands are becoming available for communication which enables applications where sensing and communication co-exist in the same band and using the same infrastructure \cite{VanderPerre2019, Behravan2022}. One proposed solution 
is RadioWeaves~\cite{VanderPerre2019} which combines distributed arrays and large intelligent surfaces \cite{Hu2018} to achieve high reliability and low-latency communication. 
Fading statistics are of great importance for the design of radio channel models and radio systems and to enable the development or investigation of e.g. network schemes and coding techniques~\cite{Oestges2011, tugratzEuCNC2023}.

Channel measurements are needed to extract the relevant parameters for realistic channel models.
In \cite{Bernado2015, Loeschenbrand2019a, Zelenbaba2021, Guevara2021, Loeschenbrand2022} measurements of distributed channels have been conducted, and for the topic of joint communication, work has mainly been done in the mmWave-bands in  \cite{PrasobhSankar2021,Zhang2022a}.
Finally, theoretical work and simulations have been performed in \cite{fascista2023uplink} for a sub-6\,GHz RadioWeave scenario for sensing.

The commonly used assumption of wide-sense stationary channels is not valid for RadioWeaves due to the large aperture.
There are two types of non-stationarities discussed in this paper. The first is related to the large aperture and the \gls{dmimo} in which the plane wave propagation assumption breaks down, i.e., operation in the near-field. Furthermore, different sub-arrays of the RadioWeave experience different channels, e.g. line-of-sight (LoS), non-LoS, or the distance to the user.
The second is the temporal non-stationarity due to the fact that the channel statistics will change over time in dynamic scenarios.

To the best of our knowledge not much effort has been devoted to RadioWeaves with very large arrays or in dynamic scenarios. 
We measure a RadioWeave with over 15000 antenna elements at sub-\SI{6}{\giga\hertz}, so that the spatial properties of the RadioWeave can be studied. 
Further, a whole new multi-link measurement system has been developed to measure the dynamic properties of the RadioWeave channels. 
With the gained insights we explore the possibilities of sub-\SI{6}{\giga\hertz} channels for joint communication and sensing.

\begin{figure}
    \centering
    \includegraphics[width=0.46\textwidth]{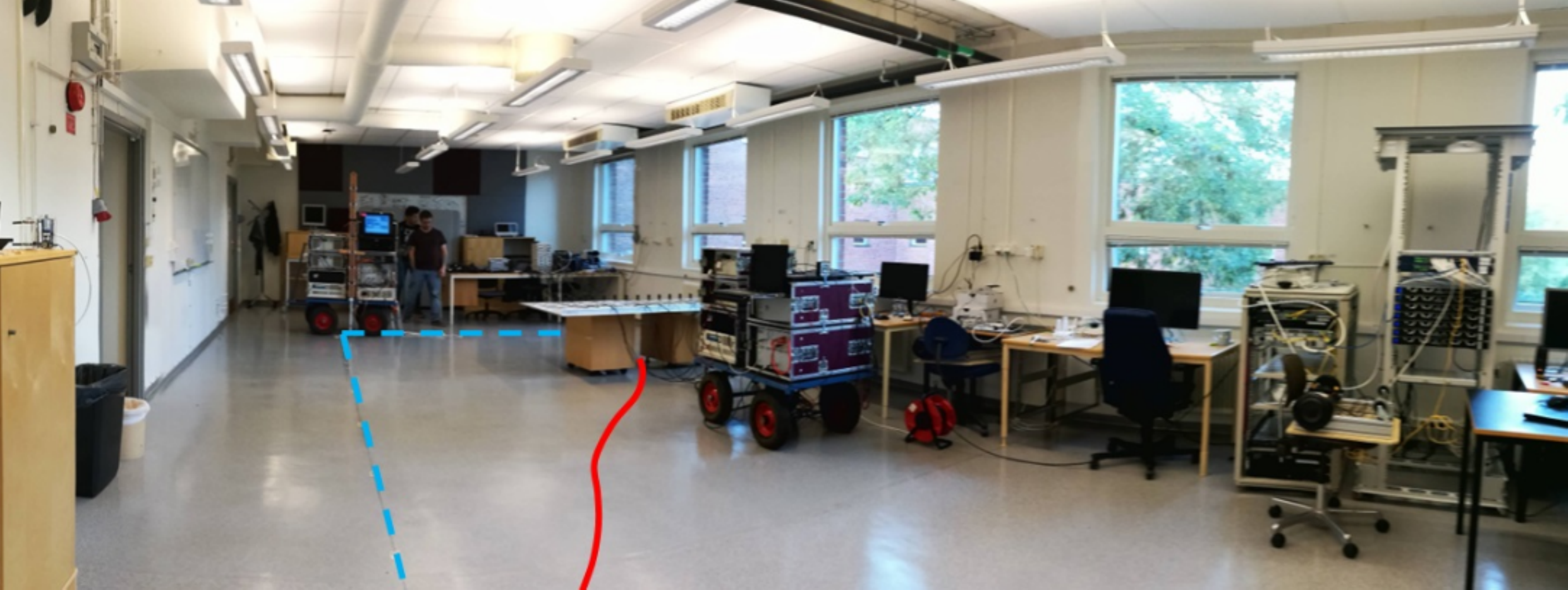}
    \caption{The indoor environment where the measurements were taken. It is a rich scattering environment with furniture or computer equipment along a majority of the walls.}
    \label{fig:5GLabPhoto}%
      \vspace{\belowFigureMargin}
\end{figure}%
\section{Measurement Campaign}
\label{sec:MeasurementCampaign}
To investigate the spatial non-stationarity due to the large antenna aperture, and the temporal non-stationarity occurring in dynamic scenarios two measurement systems were used. The systems are; 1) The RUSK LUND channel sounder for synthetic array measurements. It is a wide-band switched array channel sounder described in ~\cite{taimoor_Thesis_2014}, 2) a new distributed multi-link channel sounder to capture the dynamics of the channel.

The new channel measurement system is built around the National Instruments (NI) \gls{usrp} hardware. More specifically, the NI-USRP 2953r platform.
The \gls{usrp} consists of two RF boards with an analog bandwidth of \SI{40}{\mega\hertz} and a tuneable carrier frequency ranging from \SIrange{1.2}{6}{\giga\hertz}. Henceforth, a \textit{radio} refers to \textit{one} of these boards -- i.e. one \gls{usrp} consists of two radios.
Each \gls{usrp} connects to an external \gls{rb} oscillator, which provides a \SI{10}{\mega\hertz} reference frequency and a 1 pulse-per-second output signal to align the snapshots in the distributed system.
Before each measurement campaign, one of the \gls{rb}-oscillators is selected as the primary and is used to synchronize the clocks on the other \gls{rb}-units.
The NI LabVIEW software framework is used on the host computers to configure the radios and to acquire and store the data. The system is based on a time-division multiple access scheme that assigns all participant radios a unique time slot to transmit. During that slot, all other radios in the setup receive and store the received complex-valued samples.

The hardware for the multi-link system is summarized in Tab.~\ref{tab:MultiLinkSystem}.

\begin{table}[!ht]
    \centering
    \caption{Hardware for the multi-link measurement system.}
    \label{tab:MultiLinkSystem}
    \begin{tabular}{ll}
        \toprule
        Hardware & Description \\
        \midrule
        7 NI-USRP 2953r \SI{40}{\mega\hertz} & The radios with RF heads \\
        3 SRS FS725 & \SI{10}{\mega\hertz} frequency standard \\
        1 SRS FS740 & \SI{10}{\mega\hertz} frequency standard with GNSS \\
        7 Host computers & Radio control and logging data \\
        1 Hoverboard & Acting as mobile user\\
        12 Patch antennas & Part of the RadioWeave\\
        1 Monopole antenna & On the mobile unit\\
        \bottomrule
    \end{tabular}%
      \vspace{\belowFigureMargin}
\end{table}

 \begin{figure*}
	\captionsetup[subfigure]{labelformat=empty} 
	\centering
	\captionsetup[subfigure]{oneside,margin={0.85cm,0cm}}
	\hspace{-2mm}\subfloat[]{\vspace{3mm}\scalebox{0.82}
		{\includegraphics{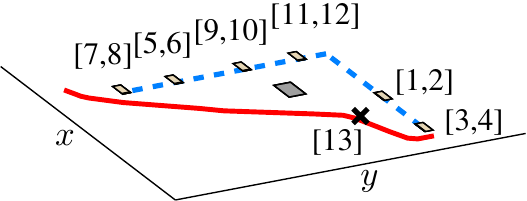}}}
	\captionsetup[subfigure]{oneside,margin={0.85cm,0cm}}
	\hspace{-19mm}\subfloat[]{\scalebox{0.86}
		{\includegraphics{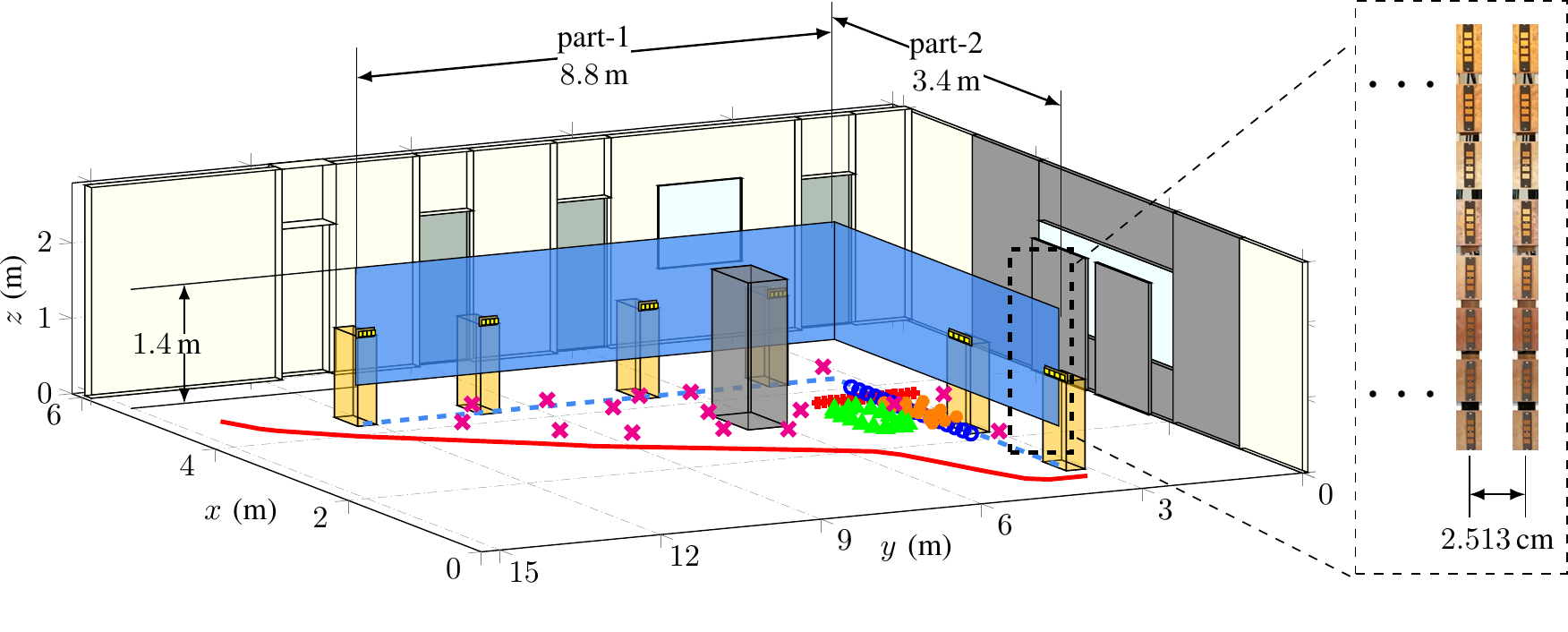}}}\\[-8mm]
	\hspace*{18mm}{\includegraphics{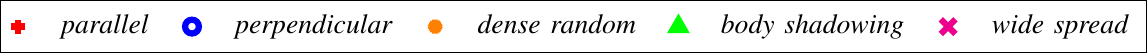}}
	\caption{3D model demonstrating the indoor environment and measurement setup. For the wide-band measurement, the blue surface denotes the synthetically constructed 3D RadioWeave and the static user locations are denoted with markers highlighted with different colors for different location setups. The gray cube represents the metal shelf filled with books. For the dynamic multi-link measurement, six two-element arrays are distributed along the 3D RadioWeave surface with the indices of the radios are shown in the left figure, and the user moving trajectory is denoted with red solid line.}
   \label{fig:MeasurementEnvironment}
   \vspace{\belowFigureMargin}
 \end{figure*}
 

The channels were measured in an indoor office-like environment at Lund University, see Fig.~\ref{fig:5GLabPhoto} and Fig.~\ref{fig:MeasurementEnvironment}. The room dimensions are approximately \SI{6 x 15 x 2.5}{\meter}.

\subsection{Wide-band Large Synthetic Array Measurement}
\label{sec:LSAScenarios}
At the transmitter side, \num{16} monopole antennas (BJTEK MA-2458) were used as \num{16} static single-antenna users. The user antenna has a quasi-omnidirectional radiation pattern in the azimuthal plane, and its vertical peak gain direction is tilted slightly upwards. At the receiver side, a 3D RadioWeave in the form of large scale array was formed. Specifically, a vertical linear array with \num{32} dual-polarized patch elements was mounted on a trolley and manually moved along a predefined L-shaped trajectory. The channel measurement from the users to the linear array was triggered by a wheel encoder after every \SI{2.513}{\centi\metre} of movement (approximately half a wavelength), therefore the horizontal spacing distance between synthetic array elements is half a wavelength. This should the fulfill the sampling theorem. As shown in Fig.~\ref{fig:MeasurementEnvironment}, the 3D RadioWeave (dimension \SI{3.4 x 8.8 x 1.5}{\metre}) consists of two mutually perpendicular synthetic planar arrays (part-\num{1} and part-\num{2}), where part-\num{1} consists of \SI{350 x 32}{} $= 11200$ patch elements and part-\num{2} consists of \SI{135 x 32}{} $= 4320 $ patch elements, so the RadioWeave contains a total of \num{15520} elements. A summary of the system parameters is listed in column \emph{Wide-band} of Tab.~\ref{tab:MultiLinkSysParams}.

Five different static user location setups were used: 1) all users closely located in a row \emph{parallel} to RadioWeave part-\num{1}; 2) all users closely located in a row \emph{perpendicular} to RadioWeave part-\num{1}; 3) all users located around a water filled human-like phantom to simulate \emph{body shadowing} effects; 4) users randomly distributed in a small area (\emph{dense random}); 5) users randomly distributed in the room (\emph{wide spread}) and a metal shelf filled with books was used as an environment shadowing object. In the first three setups, LoS propagation conditions apply to all users to the RadioWeave.

\subsection{Dynamic Multi-link Measurement}
\label{sec:MultiLinkScenarios}
The antennas described in \ref{sec:LSAScenarios} were also used during the dynamic multi-link measurements. That is, at the transmitter side,  directive patch antennas were used along its sides, and at the user side a single omnidirectional monopole was used.
The system parameters used in the measurement campaign are summarized in column \emph{Dynamic} of Tab.~\ref{tab:MultiLinkSysParams}.

A Zadoff-Chu sequence of length 449 was used for pilot signals. When a radio is active it was set to transmit \num{4} repetitions of the signal. The first one can be used as a cyclic prefix and the rest can be used for averaging to increase the \gls{snr}.

Four different scenarios were measured to investigate the channel properties for the environment depicted in Fig.~\ref{fig:5GLabPhoto}. The scenarios were: 1) the user moving along a trajectory \emph{with} a shelf in the environment, 2) the user moving along the same trajectory \emph{without} a shelf in the environment, 3) a person walking along a trajectory \emph{with} a shelf in the environment, and 4) a person walking along a trajectory \emph{without} a shelf in the environment. The user was moving along a given trajectory, c.f. red line in Fig.~\ref{fig:MeasurementEnvironment} and Fig.~\ref{fig:5GLabPhoto}. In scenarios 1 and 2 the user's direction and speed were manually controlled with a remote control (over Bluetooth), and the speed was limited to \SI{1}{\metre\per\second}.

Due to unavailable RF hardware the multi-link measurement data have not been back to back calibrated. The analysis in this paper still holds but all the power levels will be denoted as \emph{Uncal. power}. And since there are no analysis done using the impulse response the uncalibrated delay will not pose a problem at this point.
\begin{table}[!ht]
	\centering
	\caption{Channel sounding parameters.}
	\label{tab:MultiLinkSysParams}
	\begin{tabular}{p{0.18\textwidth}p{0.11\textwidth}p{0.11\textwidth}} 
		\toprule
		Description& Dynamic & Wide-band \\
		\midrule
        RadioWeave antennas         & \num{12}                      & \num{15520}\\
        user antennas               & \num{1}                       & \num{16}\\
		carrier frequency           & \SI{5.6}{\giga\hertz}        & \SI{5.6}{\giga\hertz}\\
		frequency spacing           & \SI{78.125}{\kilo\hertz}     & \SI{625}{\kilo\hertz} \\
		bandwidth                   & \SI{35}{\mega\hertz}         & \SI{240}{\mega\hertz}\\
		signal length               & \SI{12.8}{\micro\second}     & \SI{1.6}{\micro\second}\\
		signal repetitions          & \num{4}                       & \num{1}\\
		snapshot length             & \SI{665.6}{\micro\second}    & N/A\\
		repetition rate             & \SI{200}{\hertz}             & N/A\\
		max. resolvable velocity    & \SI{5}{\metre\per\second}    & N/A\\
		transmit power              & \SI{11}{\dBm}                & \SI{27}{\dBm}\\
		\bottomrule
	\end{tabular}%
   \vspace{\belowFigureMargin}
\end{table}

\section{Analysis and Results}
\label{sec:Results}

\subsubsection{Non-stationary Power Distribution}
\begin{figure}[!ht]
	\centering
    \scalebox{1}{\includegraphics{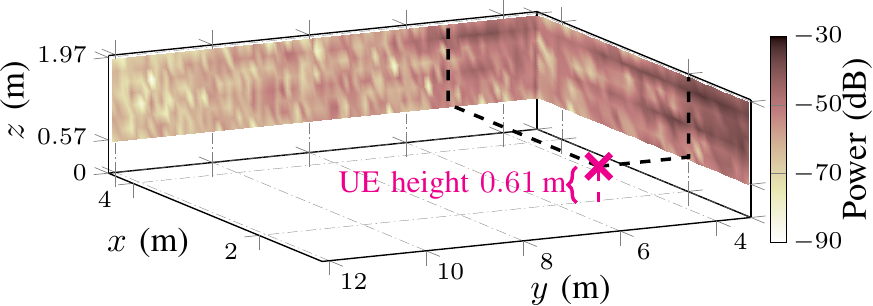}}\\[0mm]
	\caption{Received power over the 3D RadioWeave given the measured signal from the $ 3$rd user of setup \emph{wide spread}. The user position is marked with a pink ``\textbf{x}''. The black dashed lines denote the projection of user coordinates on the RadioWeave.}	 
    \label{LUfig:PD_Setup5User3}%
      \vspace{\belowFigureMargin}
\end{figure}

\begin{figure}[!ht]
    \centering
    \includegraphics{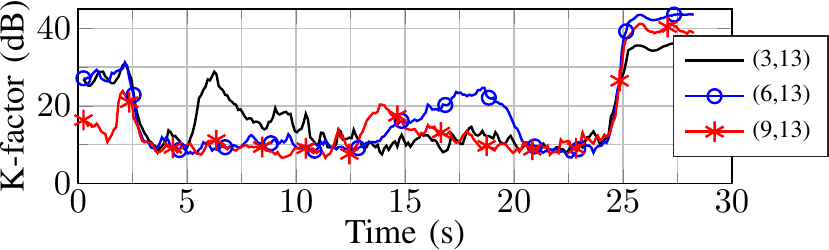}
    \caption{Examples of variations of small scale fading statistics in the dynamic case. The CDFs show the fading statistics in the three encircled regions. Here the legend describes the link between (Tx, Rx) with indices from Fig.~\ref{fig:MeasurementEnvironment}}
    \label{fig:SSF_Dynamic}%
    \vspace{\belowFigureMargin}
\end{figure}

\begin{figure}[!ht]
	\centering
	\scalebox{1}{
	\includegraphics{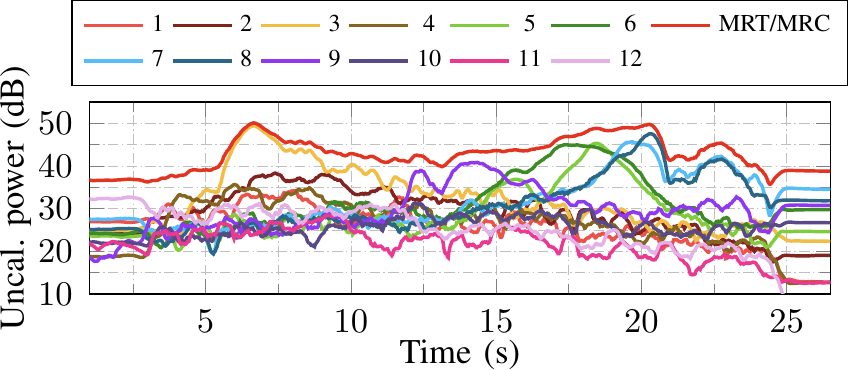}}
	\caption{Small-scale averaged received power at the user -- radio 13 -- moving along the trajectory, and a shelf as an obstructing object.}	 
	\label{fig:dynamicSsaPowerUEWithObstacle}%
   \vspace*{\belowFigureMargin}
\end{figure}

Fig.~\ref{LUfig:PD_Setup5User3} shows the power distribution across the 3D RadioWeave based on the measured signals from the $ 3$rd and $ 15$th users of the \emph{wide spread} setup.
High power intensity is observed on the sub-arrays close to the users and a maximum power variation of \SI{60}{\dB} is observed across the RadioWeave.
The non-uniform power distribution is due to the very large dimension of the array and the distances between the \gls{bs}, users, and the scatterers are smaller than the Rayleigh distance, i.e., near-field propagation.
Different sub-arrays experience different propagation environments owing to differences in geometry and antenna radiation patterns.
Moreover, propagation paths may only be visible for a portion of the array due to shadowing from objects in the environment or a human body.
Therefore, the received power from each user may vary significantly across the array elements. The spatial non-stationary property can potentially be exploited to reduce the computational complexity of algorithms, e.g., parametric channel estimation or localization, by processing only the measurement data from power-concentrated sub-arrays. 
In the dynamic scenario the temporal non-stationarity is clearly illustrated in Fig.~\ref{fig:SSF_Dynamic} where the estimated K-factor of the Ricean distribution is plotted as a function of time.
It should be noted that the high estimated K-factor in the beginning and in the end is due to the estimator since at those moments the user was static and the ratio between the dominant component and the \glspl{mpc} becomes large.
Further, in Fig.~\ref{fig:SSF_Dynamic} it is a clear that there is dominant (power) component at around 6\,seconds for link 3 which is closes to the user in the beginning. Then, as the user travels along the trajectory and the Euclidean distance to radio unit 6 decreases which results in an increasing K-factor from approximately \SI{15}{\second}.%
A typical realization of the small-scale averaged power is shown in Fig.~\ref{fig:dynamicSsaPowerUEWithObstacle}. The results are from the (robot) user moving along the trajectory with a shelf obstructing parts of the RadioWeave. It clearly shows how the received power is proportional to the LoS distance between the user and each radio unit in the RadioWeave. It also agrees with Fig.~\ref{fig:SSF_Dynamic} in the sense that when the power received from radio unit three, the K-factor is also at its highest.
%
\subsubsection{Fading Statistics}
\label{sec:Fading}
Rapid fluctuation of received amplitude within a few wavelengths is called \gls{ssf}. The fluctuation is caused by the superposition of different \glspl{mpc}. The small-scale averaged amplitude over each small spatial area typically varies over a larger spatial scale (tens to a few hundreds of wavelengths) due to shadowing by environmental objects or human bodies, i.e., \gls{lsf}~\cite{Molisch2012}. In this section, we analyze the \gls{ssf} and \gls{lsf} statistics by collectively analyzing the measurements of all \num{16} users in the \emph{wide spread} and \emph{body shadowing} setups, see Fig.~\ref{fig:MeasurementEnvironment}. For the dynamic measurements, the \gls{ssf} statistics are investigated within time frames where the channel can be approximated as stationary.

\begin{figure}[bh]
	\centering
    {\hspace*{6mm}\includegraphics{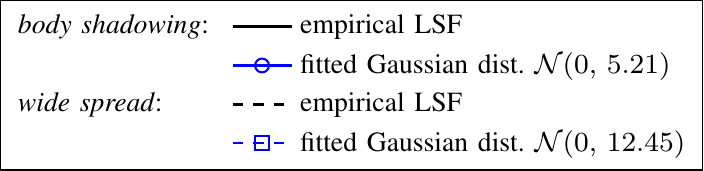}}\\[1mm]
    \includegraphics{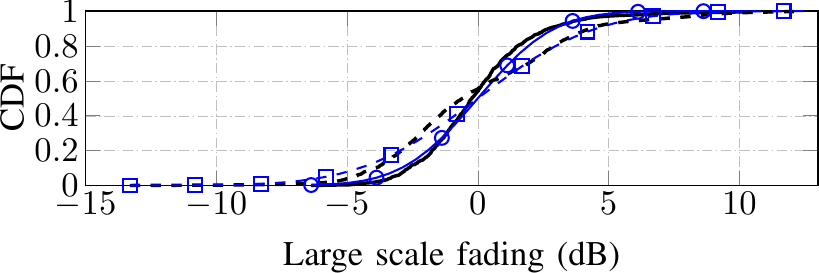}
	\caption{Large-scale fading statistics for the \emph{wide spread} and \emph{body shadowing} measurements.}	 
	\label{fig:LSF_CDF}%
   \vspace*{\belowFigureMargin}
\end{figure}

For the \emph{wide spread} setup where the users are widely distributed in the room -- larger variations of the distance and distance-dependent path loss are observed, and the path loss exponent is estimated to be \num{2.5} using a linear unbiased estimator. 
In the \emph{body shadowing} setup, where the users are in very close proximity to the water-filled phantom the transmit signal is largely shadowed leading to an estimated path loss exponent of \num{3.3}.
Fig.~\ref{fig:LSF_CDF} shows the \glspl{cdf} of the large scale fading. It can be seen that the \glspl{lsf} is well described by Gaussian distributions, but the difference between the fading statistics due to body shadowing and environment object shadowing is notable. 

\begin{figure}[!bh]
	\captionsetup[subfigure]{labelformat=empty} 
	\centering
	\captionsetup[subfigure]{oneside,margin={0.85cm,0cm}}
	\hspace{-4mm}\subfloat[]{\scalebox{0.9}
		{\includegraphics{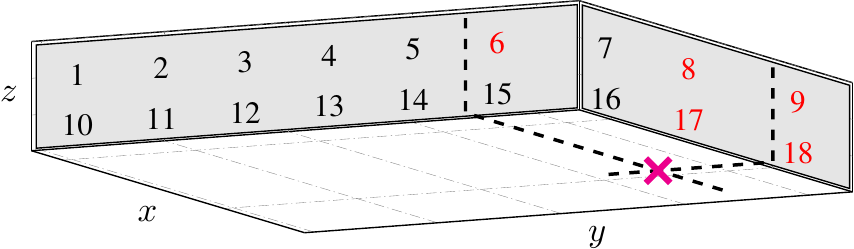}}\label{subfig:SSF_Setup5User3}}\\[-4mm]
	\captionsetup[subfigure]{oneside,margin={0.85cm,0cm}}
	\hspace{-2mm}\subfloat[]{\scalebox{0.9}
		{\includegraphics{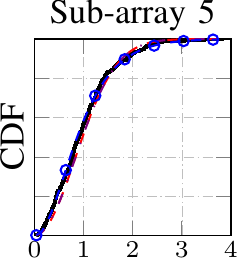}}\label{subfig:SSF_Setup5User3_seg5}}
	\captionsetup[subfigure]{oneside,margin={0.85cm,0cm}}
	\hspace{-1mm}\subfloat[]{\scalebox{0.9}
		{\includegraphics{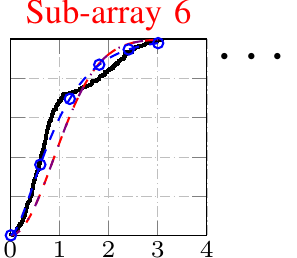}}\label{subfig:SSF_Setup5User3_seg6}}	
	\captionsetup[subfigure]{oneside,margin={0.85cm,0cm}}
	\hspace{-1.7mm}\subfloat[]{\scalebox{0.9}
		{\includegraphics{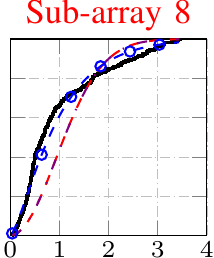}}\label{subfig:SSF_Setup5User3_seg8}}	
	\captionsetup[subfigure]{oneside,margin={0.85cm,0cm}}
	\hspace{-1mm}\subfloat[]{\scalebox{0.9}
		{\includegraphics{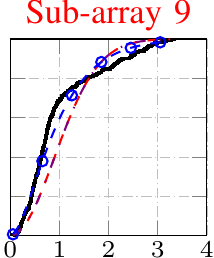}}\label{subfig:SSF_Setup5User3_seg9}}\\[-4mm]
	\captionsetup[subfigure]{oneside,margin={0.85cm,0cm}}
	\hspace{-2mm}\subfloat[]{\scalebox{0.9}
		{\includegraphics{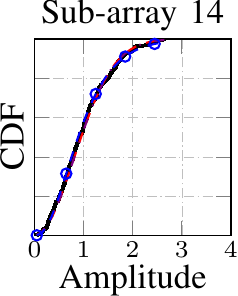}}\label{subfig:SSF_Setup5User3_seg14}}
	\captionsetup[subfigure]{oneside,margin={0.85cm,0cm}}
	\hspace{-1mm}\subfloat[]{\scalebox{0.9}
		{\includegraphics{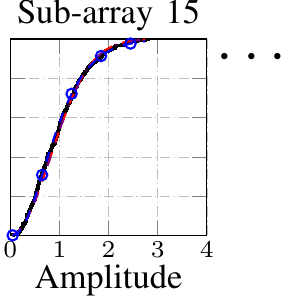}}\label{subfig:SSF_Setup5User3_seg15}}	
	\captionsetup[subfigure]{oneside,margin={0.85cm,0cm}}
	\hspace{-1.7mm}\subfloat[]{\scalebox{0.9}
		{\includegraphics{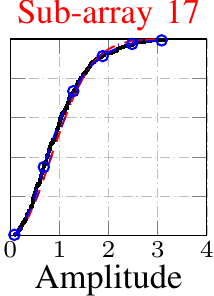}}\label{subfig:SSF_Setup5User3_seg17}}	
	\captionsetup[subfigure]{oneside,margin={0.85cm,0cm}}
	\hspace{-1mm}\subfloat[]{\scalebox{0.9}
		{\includegraphics{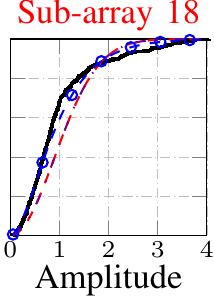}}\label{subfig:SSF_Setup5User3_seg18}}\\[-2mm]
	\hspace*{0mm}\includegraphics{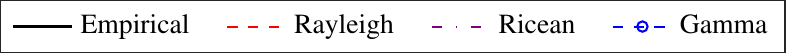}\\[-1mm]
	\caption{Small scale fading statistics of different array segments for the $3$rd user of setup \emph{wide spread}.}	 
	\label{fig:SSF_Setup5User3}%
   \vspace*{\belowFigureMargin}
\end{figure}

We further study the small-scale amplitude statistics. To get insights into the non-stationary property of the \gls{ssf} over the RadioWeave it is segmented into \num{18} sub-arrays, each with a physical size of \SI{1.3 x 1.2}{\metre}.
The empirical \glspl{cdf} of small-scale amplitudes for each sub-array and their fitted statistical distributions (Rayleigh, Ricean and Gamma) are shown in Fig.~\ref{fig:SSF_Setup5User3} for the $ 3$rd user in the \emph{wide spread} setup.

According to the non-stationary power distribution shown in Fig.~\ref{LUfig:PD_Setup5User3}, the indexes of power concentrated sub-arrays are highlighted in red.
For sub-arrays that are distant from the user, small-scale amplitudes are well described by a Rayleigh distribution or a Ricean distribution with a small K-factor.
The lack of a strong Ricean channel is contrary to the fact that a \gls{los} propagation condition applies to most of the sub-arrays. Given the presence of many highly reflective scatterers in the small measurement environment, most of the specular \glspl{mpc} are compact in the delay domain and have amplitudes comparable to that of the \gls{los} path.
Further, the antenna radiation pattern of the user leads to a low antenna gain in the direction of some sub-arrays.
For these power-concentrated sub-arrays, small-scale amplitudes are poorly characterized by either Rayleigh or Ricean distributions, while the Gamma distribution yields a better fit.

%

\subsubsection{User Separability} 
\label{sec:UserSep}
We investigate the potential ability of the RadioWeave to spatially separate proximate users. We use the channel condition number as the indicator of the degree of mutual orthogonality among user signals~\cite{Pepe_Access2018}. For the setups \emph{parallel} and \emph{perpendicular}, we select four equally-spaced users and the measurement matrix is given as $ \V{H}_{s} = [\V{h}_{s,1},\dots, \V{h}_{s,U}] $ where $ \V{h}_{s,u} $ denotes the normalized channel measurements of the $ u$th user and $ s$th setup. The channel condition number $ \kappa $ is obtained as the ratio between the maximum- and minimum singular values of $\V{H}_{s}$, i.e., $\kappa = \dfrac{\sigma_{\mathrm{max}}}{\sigma_{\mathrm{min}}}$, and $ \kappa \in [1,\infty) $. A small value close to \num{1} indicates a better spatial separability and higher achievable sum-rate given a fixed transmit power \cite{marzetta_book2016}. 

\begin{figure}[!ht]
	\centering
 	\hspace*{5mm}\includegraphics{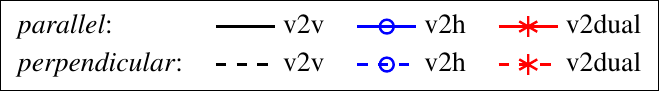}\\[2mm]
	\scalebox{1}{
		\includegraphics{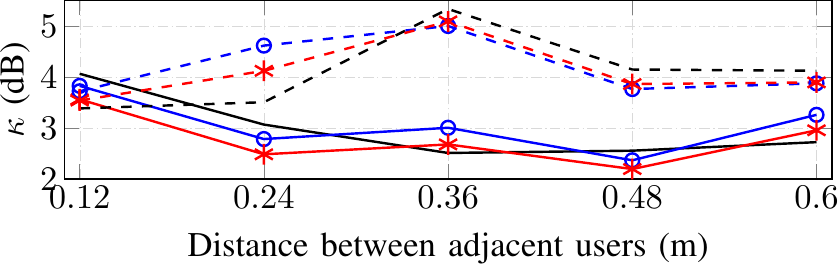}}%
	\caption{Channel condition numbers. The notations $\{\text{v2v}, \text{v2h}, \text{v2dual}\} $ denote different polarimetric transmission from the user to array elements, e.g., $ \text{v2h} $ represents vertical-to-horizontal transmission, and $ \text{v2dual} $ represents vertical-to-(horizontal \& vertical transmission).}	 
	\label{fig:CCN}%
   \vspace*{\belowFigureMargin}
\end{figure}

Previous work \cite{Pepe_Access2018} with real channel measurements have shown that massive \gls{mimo} may provide unsatisfactory performance on the user spatial separation if the channels to different users are highly correlated, e.g., in \gls{los} conditions. However, when the array size is increased to the level of RadioWeave, spatial separability can be greatly improved in these challenging situations. Fig.~\ref{fig:CCN} shows the channel condition numbers given different user spacing distances $\{0.12, 0.24, 0.36, 0.48, 0.6\}\,$\SI{}{\metre} and different polarimetric transmission schemes. In general, the result demonstrates excellent user spatial separability of the 3D RadioWeave even for the smallest user spacing. This is enabled by the large array aperture which provides high spatial resolution and the spreading scatters which help to de-correlate channels to different users.

The channel condition number for the \emph{parallel} setup is smaller than that of the \emph{perpendicular} setup, which can be explained by users of the \emph{parallel} setup benefiting more from part-1 of the RadioWeave due to its larger aperture compared with the aperture of the \emph{perpendicular} setup from part-2. Moreover, we notice that different polarimetric transmissions make no major difference on the separability.

\subsubsection{Maximum Ratio Transmission}
\label{sec:MRT}
Wireless power transmission is one of the promising applications of RadioWeaves. 
The large array aperture of the RadioWeave helps to form very narrow beamwidth with a minimum extent of $\lambda/2$ to the targeted users \cite{VanderPerre2019}, i.e., maximize the power density at the target user and minimize power dispersion to nearby users. 

\begin{figure}[!ht]
	\centering
	\captionsetup[subfigure]{oneside,margin={0.85cm,0cm}}
	\hspace{-5mm}\subfloat[\emph{parallel}, 2D]{\scalebox{1}{\hspace*{5mm}\includegraphics{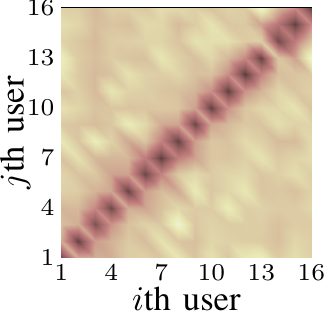}}\label{LUsubfig:MRT2D1}}
	\captionsetup[subfigure]{oneside,margin={0.85cm,0cm}}
	\hspace{-10mm}\subfloat[\emph{perpendicular}, 2D]{\scalebox{1}{\hspace*{10mm}\includegraphics{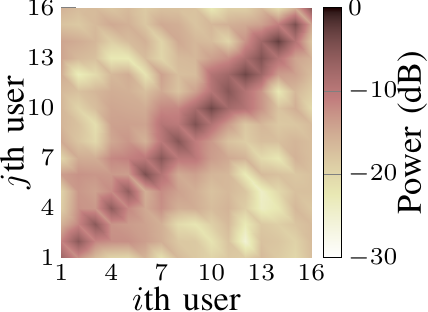}}\label{LUsubfig:MRT2D2}}\\[2mm]
	\captionsetup[subfigure]{oneside,margin={0.85cm,0cm}}
	\hspace{-5mm}\subfloat[\emph{parallel}, 3D]{\scalebox{1}{\hspace*{5mm}\includegraphics{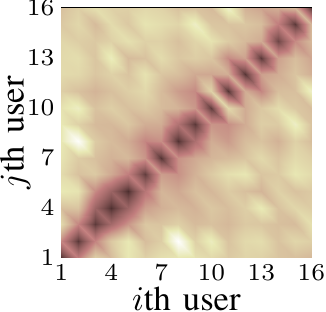}}\label{LUsubfig:MRT3D1}}
	\captionsetup[subfigure]{oneside,margin={0.85cm,0cm}}
	\hspace{-10mm}\subfloat[\emph{perpendicular}, 3D]{\scalebox{1}{\hspace*{10mm}\includegraphics{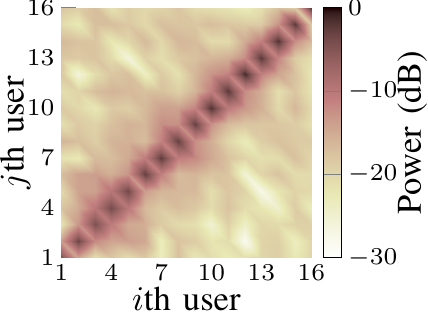}}\label{LUsubfig:MRT3D2}}\\[0mm]
	\caption{Power leakage with maximum ratio transmission to neighboring users with the 16 user setups \emph{parallel} and \emph{perpendicular}.}	 
	\label{LUfig:MRT}%
   \vspace*{\belowFigureMargin}
\end{figure}

Maximum ratio transmission \cite{MRT1999}, which maximizes the \gls{snr} at the user side, is used to perform a preliminary evaluation of the wireless power transmission capability and the interference suppression capabilities of the RadioWeave.
Fig.~\ref{LUfig:MRT} shows the leakage of power to neighboring antennas for the \emph{parallel} and \emph{perpendicular} user setups and 2D RadioWeave (part-1 only) and 3D RadioWeave (part-1 and part-2).
The amplitude of the plots in Fig.~\ref{LUfig:MRT} can be approximately interpreted as the power density at the $ j$th user's position when performing maximum ratio transmission to the $ i$th user.
Intuitively, the higher the concentration of power on the diagonal, the better the power transfer capability and the better the possibility to limit interference in the communication case.
When 2D RadioWeave is used as shown in Fig.~\ref{LUsubfig:MRT2D1} and Fig.~\ref{LUsubfig:MRT2D2}, the 2D RadioWeave shows better capability for focusing power to the target user for the \emph{parallel} setup than the \emph{perpendicular} setup, which is attributable to the large array aperture along the $ y $-axis and subsequently narrow beamwidth in the respective angular domain.
The 3D RadioWeave further extends the array aperture along the $ x $-axis which leads to  major improvement in the \emph{perpendicular} setup, as shown in Fig.~\ref{LUsubfig:MRT2D2} and Fig.~\ref{LUsubfig:MRT3D2}. 

\subsubsection{Sensing Analysis}
At \SI{5.6}{\giga\hertz} and a max relative velocity of \SI{1}{\meter\per\second} we expect to see Doppler shifts up to \SI{20}{\hertz}.
Local Fourier transforms in the time domain are performed for all the subcarriers, the ensemble mean is then taken, providing the Doppler spectrum at each time index.

\begin{figure}[!ht]
    \centering
    \includegraphics{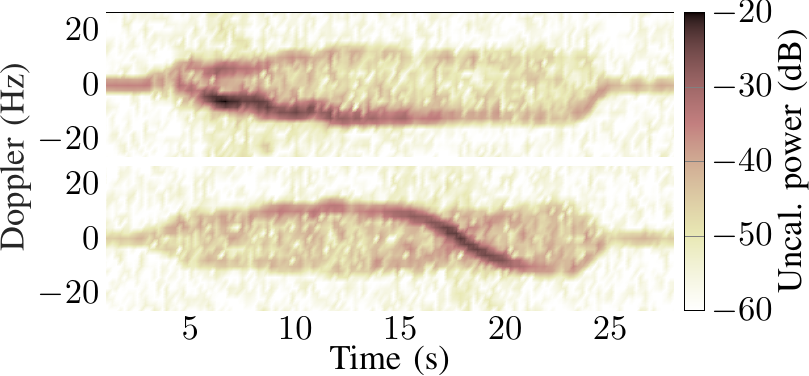}
    \caption{Time-varying Doppler spectrum for the active sensing scenario; with the user moving along the trajectory and with a shelf obstructing parts of the RadioWeave. Top: link (3,13), and bottom: link (6, 13); c.f. lower left of Fig.~\ref{fig:MeasurementEnvironment} for radio locations.}
    \label{fig:dopplerVStimeActiveSenseUE}%
      \vspace*{\belowFigureMargin}
\end{figure}

In Fig.~\ref{fig:dopplerVStimeActiveSenseUE} the time-varying Doppler spectrum is depicted for the active sensing scenario -- when the (robot) user, moved along the predefined trajectory. The top plot shows the user moving away from radio element 3 resulting in negative Doppler components. Further along the trajectory the user approaches radio element 8 to then move away. The movement is visible in the bottom of Fig.~\ref{fig:dopplerVStimeActiveSenseUE} where the Doppler crosses zero.

\begin{figure}
    \centering
 	\includegraphics{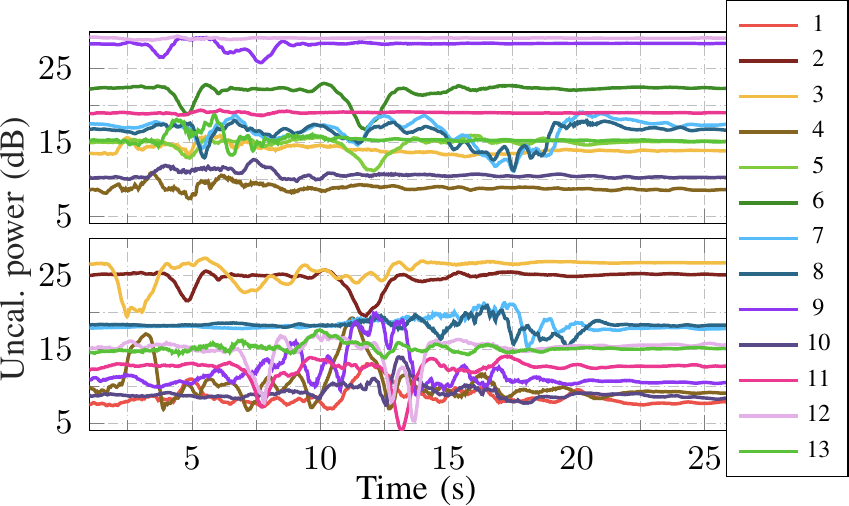}
    \caption{Small-scale averaged received power in the passive sensing scenario (multi-static radar) with a person walking  along the trajectory. Top: received power at radio element 2 in the RadioWeave. Bottom: received power at radio element 6 in the RadioWeave.}
    \label{fig:SSA_Sensing}
      \vspace*{\belowFigureMargin}
\end{figure}

For clarity -- in Fig.~\ref{fig:SSA_Sensing} we show the small-scale averaged power instead of the instantaneous power for the passive sensing scenario (multi-static radar) when a person was walking along the trajectory.
Different power variation patterns are observed for all the links at the two radio unit positions.
Two of the links of radio unit 6 are selected for additional processing.

\begin{figure}[!ht]
    \centering
    \includegraphics{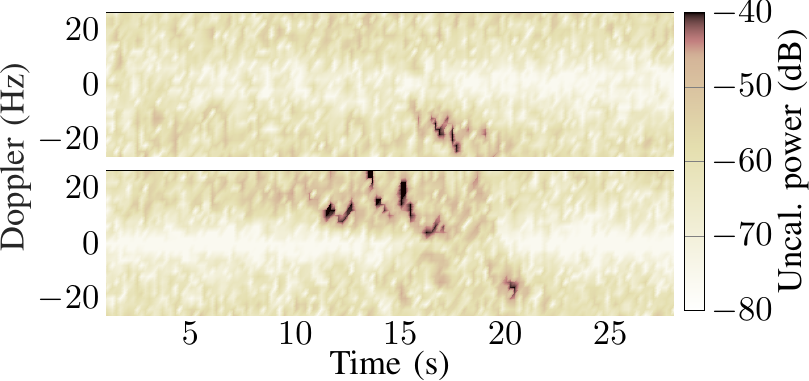}
    \caption{Time-varying Doppler spectrum for the passive sensing scenario -- with a person walking along the trajectory. Top: Link (3,6), and bottom: link (8,6); ; c.f. lower left of Fig.~\ref{fig:MeasurementEnvironment} for radio locations.}
    \label{fig:dopplerVStimeSenseRW6}%
    \vspace*{\belowFigureMargin}
\end{figure}

The results are shown for links (3,6) and (8,6) in Fig.~\ref{fig:dopplerVStimeSenseRW6}.
Even though the user walked at approximately \SI{1}{\meter\per\second} there are Doppler shifts in excess of \SI{20}{\hertz}. 
Given the evolution of the Doppler frequency in time, these contributions most likely come from the arms swinging back and forth, indicating the potential of using RadioWeave for activity detection based on the received micro-Doppler profile.

\section{Conclusions}
\label{sec:Conclusions}
This paper presented two RadioWeave measurement campaigns performed in a rich scattering indoor environment.
We used different setups in terms of array configurations, static/dynamic user behaviors, LoS, obstructing propagation conditions, and signal bandwidth to simulate different RadioWeave deployment scenarios. In particular, we introduced a new multi-link channel sounder that can capture channel dynamics in real-time. Based on the measurements, we demonstrated the deterministic and statistical non-stationary characteristics of the RadioWeave channel. We show the importance of taking non-stationarities into account – both spatial and temporal – by splitting the large RadioWeave into sub-arrays for which the statistics are stationary. The temporal and spatial power variations over the RadioWeave are significant. The non-stationary properties can potentially be exploited to reduce the computational complexity by processing only the most relevant sub-arrays, e.g., measurements from power-concentrated sub-arrays. In addition, we demonstrated the great potential of using RadioWeave for spatial separating closely located users, wireless power transfer, and sensing. 

Joint communication and sensing is a promising concept for beyond 5G communication standards, and our results suggest that the sub-6 GHz channels offer several paths for future work. Machine learning methods offer promise, as the inter-RadioWeave links experience different channels with different information that could be used for positioning. Another approach could be to use the information available in the Doppler spectra in combination with the signal strength.

\section*{Acknowledgements}
\label{sec:Acknowledgements}
The project has received funding from the European Union’s Horizon 2020 research and innovation program under grant agreement No 101013425 (Project “REINDEER”).
Parts were also supported by the Excellence Center at Link{\"o}ping–Lund in Information Technology (ELLIIT).
\bibliographystyle{IEEEtran}
\bibliography{IEEEabrv,ms}

\end{document}